\documentclass[11pt]{article} 
\usepackage[utf8]{inputenc}
\usepackage{amsmath}
\usepackage{amssymb}
\usepackage{comment}
\usepackage{float}

\usepackage{caption}
\usepackage{xcolor}
\definecolor{dgreen}{rgb}{0,0.5,0}
\definecolor{dpink}{rgb}{1,0.3,0.3}
\definecolor{darkblue}{rgb}{0,0,0.6}
\definecolor{purple}{rgb}{0.4,.2,0.7}
\usepackage[margin = 2.5cm]{geometry}
\usepackage{graphicx}
\usepackage[hyperfootnotes = false, colorlinks = true, linkcolor = blue, citecolor = purple]{hyperref}
\usepackage{subcaption}

\usepackage{cleveref}

\usepackage{fancyhdr}
\parskip=\medskipamount
\newgeometry{margin=2cm}

\def\la{\label}
\def\nref#1{(\ref{#1})}

	\newcommand{\eqn}[1]{\begin{equation}\begin{split} #1 \end{split}\end{equation}}
	\newcommand{\lp}{\left (}
	\newcommand{\rp}{\right )}
	\newcommand{\lb}{\left [}
	\newcommand{\rb}{\right ]}
	\newcommand{\RA}{\Rightarrow}

	\newcommand{\pd}{\partial}

	\newcommand{\hf}{\frac{1}{2}}
	
	\newcommand{\qrt}{\frac{1}{4}}
	\newcommand*\diff{\mathop{}\!\mathrm{d}}
	
	\usepackage{physics}
	\renewcommand{\ev}[1]{\langle #1 \rangle} %
	\newcommand{\evv}[2]{\mel{#2}{ #1 }{#2}}
	\def\i{\mathrm{i}}
	\def\t{\mathrm{T}}

\def\eep#1{ \left( #1 \right)_{E,E'} }
\def\eeptr#1{ \left( \Tr #1 \right)_{E,E'} }
\def\eepn#1{ {#1}_{E,E'} }
\def\co{\mathcal{O}}
\def\eps{\Delta E}
\def\ce{\mathcal{E}}

\begin{document}

\thispagestyle{empty}
\begin{center}
    ~\vspace{5mm}

  {\LARGE \bf {Bootstrap bounds on D0-brane quantum mechanics}}

   \vspace{0.5in}
     
   {\bf Henry W. Lin}

    \vspace{0.5in}

   ~
   \\
    Stanford Institute for Theoretical Physics, \\Stanford University, Stanford, CA 94305

    \vspace{0.5in}

    \vspace{0.5in}
    
\end{center}

\vspace{0.5in}

\begin{abstract}
We derive simple bootstrap bounds on correlation functions of the BFSS matrix theory/D0-brane quantum mechanics. The result strengthens and extends Polchinski's virial theorem bound to finite energies and gives the first non-trivial bound on $\langle{\text{Tr}\, X^2\rangle}$. Despite their simplicity, the bounds hint at some features of the dual black hole geometry. Our best lower bounds are already a factor of $\sim 2$ from  existing Monte Carlo results. 

\end{abstract}

\vspace{1in}

\pagebreak

\setcounter{tocdepth}{3}

\section{Introduction}
Solving the dynamics of strongly coupled, quantum many body systems is an arduous task.  
Strong coupling means that analytic methods are few and far between. Large $N$ makes numerics hard. 

One way of making progress is Monte Carlo, but its implementation is highly non-trivial. First, large $N$ extrapolation is computationally expensive, especially when the number of lattice sites is large as required for large inverse temperatures. Second, Monte Carlo is best suited for problems in Euclidean signature where the measure is real and positive. This means that Lorentzian setups, as well as Euclidean problems with fermions, must be approached with caution. Finally, some models at finite $N$ are only metastable.

Although technically daunting, the reward for solving such large $N$, chaotic quantum systems promises to be hefty, in two senses of the word. The chief example is the D0-brane quantum mechanics \cite{deWit:1988wri}, or BFSS matrix theory \cite{Banks:1996vh}. In the t' Hooft limit, this model is holographically dual to a certain black hole \cite{Itzhaki:1998dd}, see \cite{Maldacena:2023acv} for a recent review. Solving the model therefore promises new insights into quantum black holes.
For the reasons mentioned above, few analytical results have been achieved on the matrix model side.  However, despite the aforementioned difficulties, heroic Monte Carlo simulations have been performed \cite{Kabat:2000zv, Anagnostopoulos:2007fw, Hanada:2008ez, Catterall:2008yz, Filev:2015hia, Kadoh:2015mka, Berkowitz:2016jlq, Berkowitz:2018qhn, Pateloudis:2022ijr}. After performing continuum/large $N$ extrapolations, the authors found remarkable agreement with the predicted black hole thermodynamics, and even gave new predictions for the $1/N$ and stringy $\alpha'$ corrections. This is particularly impressive given both the fermion sign problem and finite $N$ metastability of the model.

A recent technique that is complementary to Monte Carlo is the matrix/quantum mechanical bootstrap \cite{Anderson:2016rcw,Lin:2020mme,Han:2020bkb,Berenstein:2021dyf,Kazakov:2021lel,Kazakov:2022xuh,Nancarrow:2022wdr}. The bootstrap is a modern incarnation of the ancient and venerable method of ``guess and check.''
For quantum mechanical problems \cite{Han:2020bkb}, one starts by ``guessing'' the expectation value of some set of simple operators $\ev{O_i}$ in some state $\rho$ that satisfies $[H, \rho]=0$. Then, the relation $\ev{[H,O_i]}  = 0$ allows us to infer the expectation values of new operators $\ev{O'_i}$ based on our guess. 
To check the guess, we consider an arbitrary $\mathcal{O}$, which is some superposition of all the operators which we know something about. If our initial guess was correct, positivity of the inner product guarantees:
\eqn{
\la{basic}
\langle \co^\dagger \co \rangle \ge 0.
}
So if we manage to find some $\co$ such that $\ev{\co^\dagger \co }$ is negative, we can rule out our initial guess. By searching over possible guesses, one obtains an allowed region in the space of correlation functions. In favorable settings, this region shrinks as we impose more and more constraints. 
In principle, one can bootstrap any quantum mechanical system, but for large $N$ matrix systems in the 't Hooft limit, we can exploit large $N$ factorization to restrict $\mathcal{O}$ to single trace observables $\mathcal{O} = \Tr ( \cdots )$. Imposing this factorization\footnote{In addition, one should consider constraints coming from the cyclicity of the trace (over the SU$(N)$ indices), canonical commutation relations and gauge constraints \cite{Han:2020bkb}.  } is the key to breaking the curse of dimensionality in the bootstrap \cite{Anderson:2016rcw, Lin:2020mme, Han:2020bkb} and also allows one to work directly in the infinite $N$ limit where metastability is tamed\footnote{See \cite{Lin:2020mme} for a bootstrap analysis of a single matrix model with unbounded matrix potential.}. Further, since the method of \cite{Han:2020bkb} only uses the positivity of the Hilbert space inner product \nref{basic}, fermions are not an issue.

A natural question is whether the bootstrap can be usefully applied to the BFSS matrix theory, which is more complicated than the models that have already been bootstrapped. In fact, this question was already answered out-of-time-order in 1999 by Polchinski \cite{Polchinski:1999br}. Polchinski provided one of the few known analytic data points on the model by constraining the moments of the bosonic matrices in the ground state. From a modern viewpoint, Polchinski's argument can be rephrased as a bootstrap bound. We will rederive this result from the bootstrap perspective; besides a mere rephrasing of the argument, the bootstrap perspective will allow us to improve the bound in various directions. In particular, we both improve the ground state bound and extend the bound to $E > 0$, as well as deriving constraints on other operators.

The bootstrap approach can be pursued both analytically and numerically. In this paper, we will only carry out a few rounds of the ``guess and check'' procedure that can be performed analytically, by judiciously choosing both what to guess and what to check. To carry out further rounds, one will presumably need numerics. By demonstrating that the first few rounds involving only simple operators already yield non-trivial information, we hope to motivate a more systematic numerical bootstrap search in the future.

A short review along with a minor generalization of the quantum mechanical bootstrap that gives information about off-diagonal elements of operators $|\bra{E'}\mathcal{O} \ket{E}|$ is presented Appendix \ref{SQM}, see also \cite{Nancarrow:2022wdr}.

\section{Bootstrapping BFSS}
\subsection{Definition of the model and review of its gravity dual}
The BFSS matrix theory consists of 9 bosonic matrices $X_I$ and 16 fermionic matrices $\psi_\alpha$, which transform under an SO(9) $R$-symmetry in the fundamental and spinor representations. All matrices are taken to be Hermitian and traceless; more explicitly, 
we can expand $\psi_\alpha = \psi_\alpha^A \t_A, X_I = X_I^A \t_A$ where $\t_A$ is a basis (over the reals) for the $N^2-1$ traceless Hermitian matrices. Then each $\psi_\alpha^A$ is just a Majorana fermion and similarly $X_I^A$ is just a non-relativistic particle with canonical commutation relations
\begin{equation}
\begin{split}
\label{canonical}
\{  \psi_\alpha^A , \psi_\beta^B \} =  \delta^{AB} \delta_{\alpha \beta} , \quad [X_I^A,P_J^B] = \i \, \delta^{AB}\delta_{IJ} . %
\end{split}
\end{equation}

The Hamiltonian is
 \begin{equation}
\begin{aligned}\la{ham}
H &= \hf  \operatorname{Tr}  \left(g^2   {P_I^2}-\frac{1}{2g^2}\left[X_I, X_J\right]^2-\psi_\alpha \gamma^I_{\alpha \beta} \left[ X_I,\psi_\beta\right] \rp .
\end{aligned}
\end{equation}
In the above expression, there is an implicit sum over $I,J$. 
With these conventions, $X$ has units of energy and $g^2$ has units of $E^{3}$.
We can take the SO(9) gamma matrices $\{ \gamma^I, \gamma^J\} = 2 \delta^{IJ}$ to be real, traceless, and symmetric.
This model has 16 supercharges which transform as spinors under the SO(9) global symmetry. The only consequence of supersymmetry that we will use is that $H \ge 0$.%

In addition, the model has an SU$(N)$ symmetry where each matrix transforms in the adjoint representation. One can choose to treat the SU$(N)$ symmetry as either a global or a gauge symmetry \cite{Maldacena:2018vsr}; the bounds  we derive will be agnostic to this choice.

We will be mostly interested in the theory in the 't Hooft limit \cite{Itzhaki:1998dd} see \cite{Polchinski:1999br, Maldacena:2018vsr, Maldacena:2023acv} for a review. This is a limit $N \to \infty$ while also taking a dimensionless coupling constant fixed. Since $\lambda = g^2 N$ is a dimensionful quantity, we need to combine it with an energy scale. If we study the theory at some temperature $T$, the effective coupling is $\lambda/T^3$. Holding this combination fixed while taking $N \to \infty$ is the 't Hooft limit, where %
the theory is dual to a charged 10D black hole in type IIA string theory. The metric and dilaton in the black hole background are given by
\def\tir{\tilde{r} }
\begin{equation}
\begin{split}
\label{metricIIA}
  {\diff s^2 \over \alpha'} &= -f(r) r_c^2 \diff t^2 + \frac{\diff  r^2}{f(r) r_c^2}+ \lp r \over r_c \rp^{-3/2} \diff \Omega_8^2\\
  f(r)&= \lp 1-\frac{r_h^7}{r^7}\rp \lp r \over r_c \rp^{7/2} \\
    e^{-\phi} &= {60 \pi^3 N}  \lp r \over r_c \rp^{21/4}, \quad r_c = (240 \pi^5 \lambda)^{1/3}. %
\end{split}
\end{equation}
The SO(9) symmetry of the quantum mechanics is realized as the isometries of the $S^8$. Note that the size of the $S^8$ shrinks with $r$, $R_\text{eff}^2/\alpha' = (r_c / r)^{3/2}$.
Eventually the effective radius of the $S_8$ is comparable to the string scale, at which point the geometry cannot be trusted. Since the horizon radius grows with temperature $r_h = \left( 4 \pi T / 7  \right)^{2/5} r_c^{3/5}$, the regime in which any portion of the supergravity solution to be valid is:
\begin{equation}
\begin{split}
\label{temp_range}
  T^3 \ll \lambda, \quad r^3 \ll \lambda.
\end{split}
\end{equation}
One can compute the entropy of the black hole $S = A/(4G_N)$ using the Bekenstein-Hawking formula in Einstein frame, with $16 \pi G_N = (2\pi)^7 (\alpha')^4$. This gives the thermodynamics:%
\begin{equation}
\begin{split}
\label{thermo}
\frac{E}{N^2} &= \lambda^{1 / 3} \frac{9}{14} 4^{13 / 5} 15^{2 / 5}(\pi / 7)^{14 / 5}\left(\frac{T}{\lambda^{1 / 3}}\right)^{14 / 5}.
\end{split}
\end{equation}
At energies $E/N^2  \gtrsim \lambda^{1/3}$, the thermodynamics is significantly modified by $\alpha'$ corrections. This equation was checked on the matrix side using Monte Carlo \cite{Kabat:2000zv, Anagnostopoulos:2007fw, Hanada:2008ez, Catterall:2008yz, Filev:2015hia, Kadoh:2015mka, Berkowitz:2016jlq, Berkowitz:2018qhn, Pateloudis:2022ijr}.

At finite $N$, all energy eigenstates besides the ground state are scattering states.  From the gravity point of view, the black holes can Hawking evaporate into D0-branes; on the matrix side there are flat directions in the potential $\Tr [X_I, X_J]^2$. Since we are dealing with scattering states, all moments $\ev{\Tr X^\ell}$ should diverge for finite energy eigenstates.\footnote{Another option for dealing with is to consider the BMN matrix theory \cite{bmn} with a small $\mu$, where the mass terms modify the large $X$ region so that there are no scattering states. Presumably the correlators can be defined by first taking $N\to \infty$ first and then taking $\mu \to 0$ (in that order). See also \cite{Catterall:2009xn, Pateloudis:2022ijr} for a discussion of the issue in Monte Carlo simulations. I thank Xi Yin for this discussion.}
Nevertheless, at large $N$, the black hole states are metastable with a lifetime $\tau$ that grows with $N$. 
This suggests that if we consider a state $\rho$ such that $[H,\rho]$ is suppressed by $N$, then $\ev{\Tr X^\ell}$ would be finite. In other words, we consider density matrices that are nearly time-independent, such that $\delta E \sim \i [H,\rho]$ is of order the inverse lifetime $\tau$ of the black hole. Since $\tau \to \infty$ at large  $N$, we can ignore this subtlety in the large $N$ limit. In practice, inputting large $N$ factorization should likely resolve this subtlety; see \cite{Lin:2014wka} for a somewhat analogous problem involving a single matrix integral that is ill-defined at finite $N$.

One question is whether simple observables such as $\ev{\Tr X^\ell}$ at finite temperature have a supergravity interpretation. The holographic dual of such observables is currently unknown, but we make some comments about thermal 1-pt functions in Appendix \ref{thermal1pt}. 
A heuristic argument \cite{Polchinski:1999br} is that $\Tr X^2$ roughly measures the size of the supergravity region, since the off-diagonal elements of $X$ roughly measure the mass of a string stretched between D-branes. Some further comments on this picture are discussed in Appendix \ref{thermal1pt}, see also \cite{Hanada:2021ipb}.

\subsection{A bootstrap bound on BFSS \la{polchinski} }

Let's choose arbitrarily one of the 9 bosonic matrices and denote it by $X$ and its conjugate momentum by $P$. We will consider expectation values of single trace operators $\ev{\Tr(\cdots)}$ in a state $\rho$ such that $\ev{ H} = E$ and $[H,\rho]$ is negligible in the large $N$ limit.
To apply the bootstrap philosophy, we imagine guessing the values of $\ev{\Tr X^4}$. We will consider several different operators $O$ and apply $[H,O]$ to generate expectation values that are related to $\ev{\Tr X^4}$. By checking positivity of the inner product, we will derive a bound on a tenable ``guess'' for this correlator.
The bounds in this section result from two rounds of bootstrap constraints. Each round consists of applying $\ev{[H,O]} = 0$ followed by positivity constraints.

{\bf Round 1.} We start with the bootstrap constraint $\ev{ [H,\Tr X^2]} = 0$, which implies $ \ev{\Tr X^I P_I + P^I X_I } =0$.
Together with $\Tr [X,P] = \i N^2$, we learn\footnote{For SU(N) matrices we actually have $\Tr [X,P] = \i (N^2-1)$ but we will drop the $-1$. Keeping track of this would yield a slightly stronger bound at finite $N$. } that $\ev{\Tr XP}=-\ev{\Tr PX} = \i N^2/2$. 
Then we apply our first positivity constraint
\eqn{\mathcal{M} = \left(
\begin{array}{cc}
\ev{\Tr X^2}  & \ev{\Tr X P} \\
\ev{\Tr  P X} & \ev{\Tr P^2} \\
\end{array}
\right) \succeq 0 \quad \RA \quad \sum_I \ev{ \operatorname{Tr} X^2 } \ev{ \operatorname{Tr}\left(P^{I} P_{I}\right) } \geq {9\over 4} N^{4}.
\la{uncertain}
}
Here the symbol $\succeq$ means that all eigenvalues are non-negative.
\nref{uncertain} is basically just the uncertainty principle\footnote{The uncertainty principle was also used by Polchinski in \cite{Polchinski:1999br}, however, a factor of 4 was neglected in equation 7.4 of  \cite{Polchinski:1999br}. }. We write the Hamiltonian $H = K + V+ F$ where $K,V$ are the kinetic and potential terms and $F$ is the fermionic term. The term that shows up in \nref{uncertain} is $K$. We want to replace with $V$, which is quartic in $X^I$ in order to make contact with $\ev{\Tr X^4}$.

{\bf Round 2.}   To do so, we use $\ev{[H, \Tr X P]} = 0$ and  $\ev{H} = E$, which gives:
\begin{equation}
\begin{split}
\label{virial_idea}
  \ev{-2K + 4 V + F} = 0, \quad \ev{K} + \ev{V} + \ev{F} = E.
\end{split}
\end{equation}
Eliminating $\ev{F}$ gives the relation 
\begin{equation}
\begin{split}
\label{kinetic_pot}
  2\ev{K} = \tfrac{2}{3}E + 2\ev{V}.
\end{split}
\end{equation}
Now we want to relate $V$ to the simpler operator $\Tr X^4$. To do so, we apply another round of positivity constraints.
First, consider the operators $X^2, Y^2, XY, YX$ (where $X$ and $Y$ are any two of the matrices $X_I$) and two matrices of correlators: %
\eqn{
\left(
\begin{array}{cc}
\ev{\Tr X^4}  & \ev{\Tr X^2 Y^2} \\
\ev{\Tr   X^2 Y^2} & \ev{\Tr Y^4} \\
\end{array}
\right) \succeq 0, \quad 
\left(
\begin{array}{cc}
\ev{\Tr X^2 Y^2}  & \ev{\Tr XYXY} \\
\ev{\Tr   XYXY} & \ev{\Tr X^2 Y^2} \\
\end{array}
\right) \succeq 0, 
}
which leads to the bound
\begin{equation}
\begin{split}
\label{ineq_comm}
  -\langle{\Tr [X,Y]^2}\rangle = 2 \ev{\Tr X^2Y^2} -  2\ev{\Tr XYXY}  \le  4  \ev{\Tr X^4}. \end{split}
\end{equation}
For an SO(9) rotationally invariant state, we conclude that $(9 \times 8 ) 4 \Tr X^4 \ge  -\sum_{m, n}\langle  \operatorname{Tr} \left[X^{m}, X^{n}\right]^{2} \rangle = 4\ev{V}$. Substituting this into \nref{uncertain} with the help of \nref{kinetic_pot}, 
  \eqn{\ev{\operatorname{Tr} X^2}   \lp {144 \over g^2}\ev{\Tr X^4}+{2 E \over 3} \rp   \geq {9\over 4} g^2 N^4 .}
Using one more positivity constraint %
\eqn{\la{Cm1} \mathcal{M}_1 = \left(
\begin{array}{cc}
\ev{\Tr 1}  & \ev{\Tr X^2} \\
\ev{\Tr  X^2} & \ev{\Tr X^4} \\
\end{array}
\right) \succeq 0,}
we find
\eqn{ \langle{\operatorname{tr} \tilde{X}^4}\rangle^{1/2}  \lp 144 \langle{\operatorname{tr} \tilde{X}^4}\rangle+ \tfrac{2}{3} \mathcal{E}  \rp \ge {9\over 4} , \quad \mathcal{E} = \lambda^{-1/3} {  E \over   N^2},\quad \tilde{X} = \lambda^{-1/3} X .
\la{Polchinski}}
Here we have rewritten the constraint using variables that are natural in the 't Hooft limit. %
We have also introduced ``little trace'' which satisfies $\tr \mathbf{1} = 1$ whereas ``big Trace'' satisfies $\Tr \mathbf{1} = N$. 
We plot this bound in {\color{red} red} in Figure \ref{fig:t4_constr}.
If we are interested in the ground state, we can set $E=0$ and obtain the estimate $\Delta X \sim \lambda^{1/3}$ found in Polchinski \cite{Polchinski:1999br}. This agrees with the size of the gravity region in the 't Hooft limit \cite{Polchinski:1999br, Itzhaki:1998dd}.

Notice that the bound changes significantly at energies $E \lambda/N^2 \sim \lambda^{4/3}$.  
We can convert this into a temperature using the gravity solution \nref{thermo}   $T \sim \lambda^{1/3}$, which is precisely the regime of validity of the gravity solution \nref{temp_range}. %
In other words, even the simplest bootstrap bounds hints at aspects of the emergent geometry. At small energies, the breakdown of the geometry near the boundary is diagnosed by the size of the matrices, whereas at finite energy the breakdown of the geometry near the horizon is related to a transition in the bound. We will see more evidence for this claim in section \ref{fermion}.

We could ask whether the above bound gives something a bound on $\ev{\tr X^2}$ and not just $\ev{\tr X^4}$. One can use the inequality $-\tr [X,Y]^2 \le 2 N \tr X^2 \tr Y^2$ \cite{bottcher2008frobenius}. This gives a lower bound on $(\tr X^2)^3$ that scales like $\lambda^2/N$ which is trivial in the 't Hooft limit.
Incorrectly assuming that this bound is parametrically saturated would give a different large $N$ scaling for the typical eigenvalues of one of the bosonic matrices. So we need additional input to get a non-trivial bound on $\ev{\tr X^2}$. %

\subsection{Constraints from the fermions \la{fermion}}
In this subsection, we will consider the bounds coming from the fermionic terms in the Hamiltonian. By considering these terms, we can get a bound on $\ev{\Tr X^2}$ which was inaccessible in \ref{polchinski}. We will also improve on the bound of $\ev{\Tr X^4}$.
In the previous section, by eliminating the fermionic term to write \nref{kinetic_pot} we only used half of the information in \nref{virial_idea}. %
We can also solve for the fermionic term:
\begin{equation}
\begin{split}
\label{fev}
  \ev{F} = 2 \lp \tfrac{1}{3} \ev{E} - \ev{V} \rp. %
\end{split}
\end{equation}
This equation becomes useful when we observe that $F$ is schematically of the form $F \sim \psi \psi X$. Since $\psi$ is a collection of Majorana fermions, $\psi \psi$ is a bounded operator, so the only way $F$ can be large is for $X$ to be large. Thus we expect a lower bound on $\ev{\Tr X^2}$.

Now let's give the precise argument. First, we may rewrite the fermionic term in \nref{ham} as
\eqn{F = \hf \gamma^I_{\alpha \beta}  \Tr  \{\psi^\alpha ,  \psi^\beta \}  X^I \equiv \Tr O_I X^I .}
 We want to assemble a matrix of correlators involving the operators $O_I, X_I, P_I$. From Section \ref{polchinski}, we got constraints on $\Tr P^2$ and $\Tr XP$.  Using \nref{canonical} and the constraint $\ev{[H,F]}= 0$ gives $  \ev{\Tr O^I P_I} = 0$, so:
 \begin{equation}
 \la{Cm2}
 \mathcal{M}_2 = \lb
 \begin{matrix}
\tfrac{1}{9} \ev{\Tr O_I O_I}  & \tfrac{2}{9} \lp \tfrac{1}{3}E-\ev{V} \rp  &  0\\
\tfrac{2}{9} \lp \tfrac{1}{3}E-\ev{V} \rp  & \ev{\Tr X^2} &\i \hf N^2 \\
0 & -\i \hf N^2  & \tfrac{2}{9} \lp \tfrac{1}{3}E+\ev{V} \rp
\end{matrix}\rb \succeq 0.
 \end{equation}

The term $\Tr O_I O_I$ just consists of Majorana fermions, so it cannot be arbitrarily large. To bound this term, consider  $I=2$. In our conventions\footnote{More generally, any of the $\gamma^I$ has half of its eigenvalues $+1$ and half of its eigenvalues $-1$. } $(\gamma^2)_{\alpha \beta} = s_\alpha \delta_{\alpha \beta}$ where $s_\alpha = +1$ for $\alpha = 1, 2, \cdots 8$ and $s_\alpha = -1$ for $\alpha = 9, 10, \cdots 16$. Therefore,
\eqn{ \la{fermionBd}
  \frac{1}{9}\ev{\Tr O_I O_I } &=  \ev{\Tr O_2 O_2} =\sum_{\alpha,\beta} s_\alpha s_\beta \ev{\Tr   \psi_{\alpha}^2 \psi_\beta^2}  \le 64 N^3 .} 
In the last line we used equation \nref{result4} in Appendix \ref{fermionMatrices}. This is a non-trivial bound since the trace gives rise to $N^4$ terms.
Now we can minimize $\ev{\Tr X^2}$  subject to the constraints \nref{ineq_comm}, \nref{Cm1}, \nref{Cm2}, and \nref{fermionBd}.
We find that the boundary of the allowed region is achieved when $\det \mathcal{M}_2 = 0$ and $\ev{\Tr O_2 O_2} \to  64 N^3$. This implies 
\begin{align}
	&\langle \tr \tilde{X}^2 \rangle \ge \frac{(\mathcal{E} - 3 v)^2}{9^3 \times 16} + \frac{27}{8(\mathcal{E} +3 v)}, \la{firstLine} \\
	&\ce^2 + \frac{3^9}{\ce + 3 v} = 9 v^2 \la{secondLine}.
\end{align}
Here $v$ is the boundary value of $\ev{V}$. We minimized the right hand side of \nref{firstLine} with respect to $v$ to obtain \nref{secondLine}.
 This constraint is plotted in Figure \ref{fig:t2}. At $E = 0$ this gives the bound $\ev{\tr \tilde{X}^2 } \ge 3/16$, whereas at large energies the bound gets weaker $\sim 27/(16\mathcal{E})$. %

We can compare the bounds that we obtained to the most recent Monte Carlo results \cite{Pateloudis:2022ijr}. These were obtained by simulating the BMN model \cite{bmn} at small mass parameter $\mu$ at various $N$ and lattice spacings (in Euclidean time). Then a large $N$/continuum extrapolation was performed to arrive at the results shown in Figure \ref{fig:t2}. Surprisingly, our lower bound is only a factor of $\sim 2$ smaller than the Monte Carlo result at the smallest temperature. We also compared with the $N=32$ results reported in Appendix B of \cite{Berkowitz:2016jlq}. We took the largest values from the largest number of lattice sites $L$ available $L = 32$ or $L=16$; we did not perform any large $L$ or large $N$ extrapolation, which presumably accounts for the small difference in the {\color{blue} blue} and {\color{red} red} points in Figure \nref{fig:t2}. %

\begin{figure}[t]
    \begin{center}
  \hspace{1cm}  \includegraphics[width=0.7\columnwidth, trim={0 0cm 0 0cm}, clip]{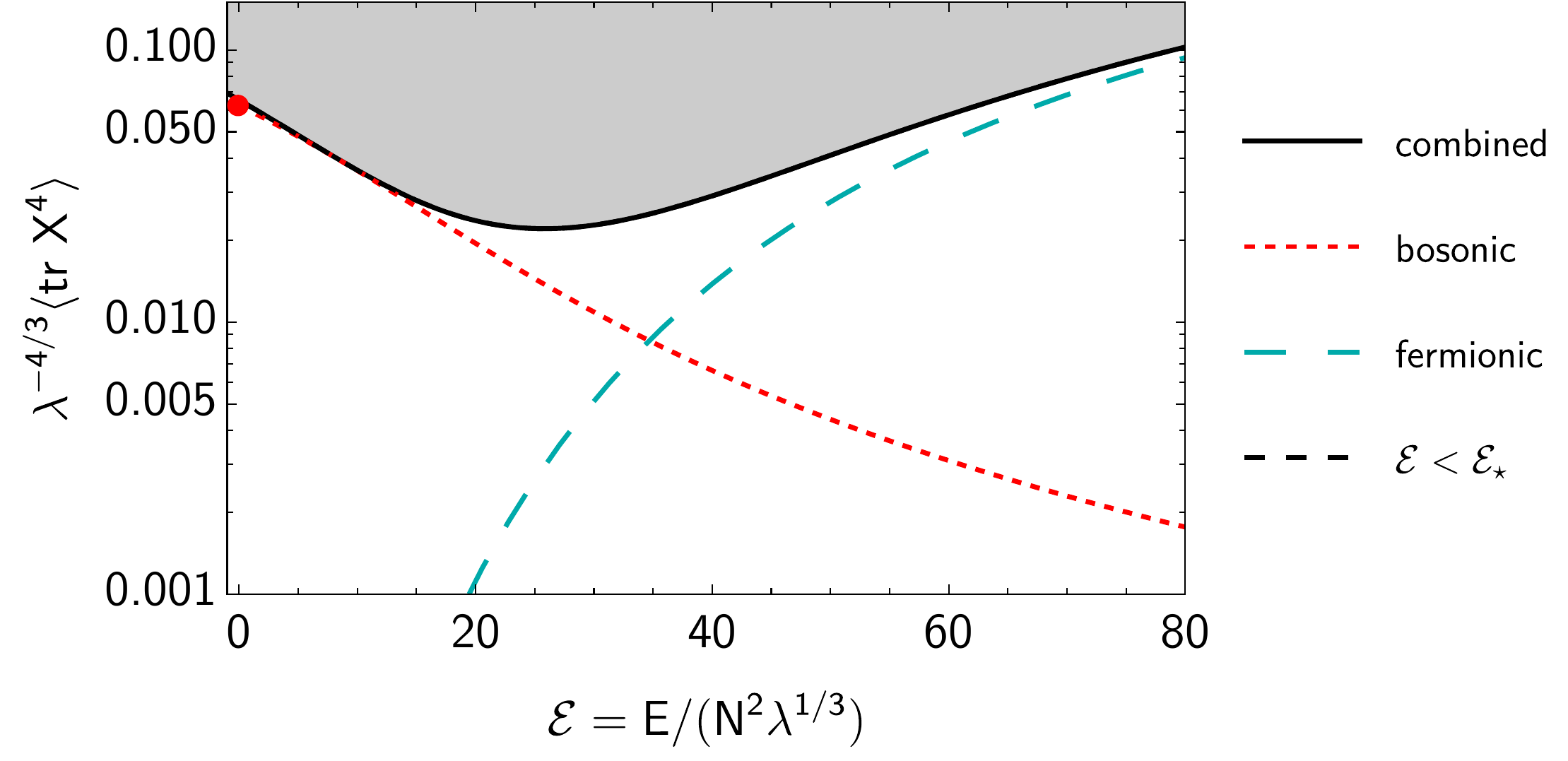}
   \vspace{-.5cm}
    \end{center}
    \caption{Lower bounds on the correlator $\ev{\tr X^4}$ as a function of energy, where $X$ is any one of the 9 bosonic matrices $X^I$. We are plotting dimensionless quantities on both axes. The ``Polchinski point'' --- indicated by a zero energy {\color{red} dot} --- was derived in \cite{Polchinski:1999br} up to factors of 2. A simple generalization of the bound is the dashed {\color{red} red curve}, see equation \nref{Polchinski}. We can improve the bound by considering combining these constraints with correlators involving fermions. The best constraint that we derived is shown in solid black. (Although it is difficult to see, the black curve lies above the red point at $E=0$.) At large energies $E/N^2 \gg \lambda^{1/3}$ the constraints primarily come from the {\color{teal} fermionic correlators}. 
     The characteristic scale on which the bound varies is $E/N^2 \sim \lambda^{1/3}$ which corresponds to the scale at which the supergravity solution is breaking down. }
    \label{fig:t4_constr}
\end{figure}

\begin{figure}[t]
      \begin{center}
  \hspace{1cm}\includegraphics[width=0.7\columnwidth]{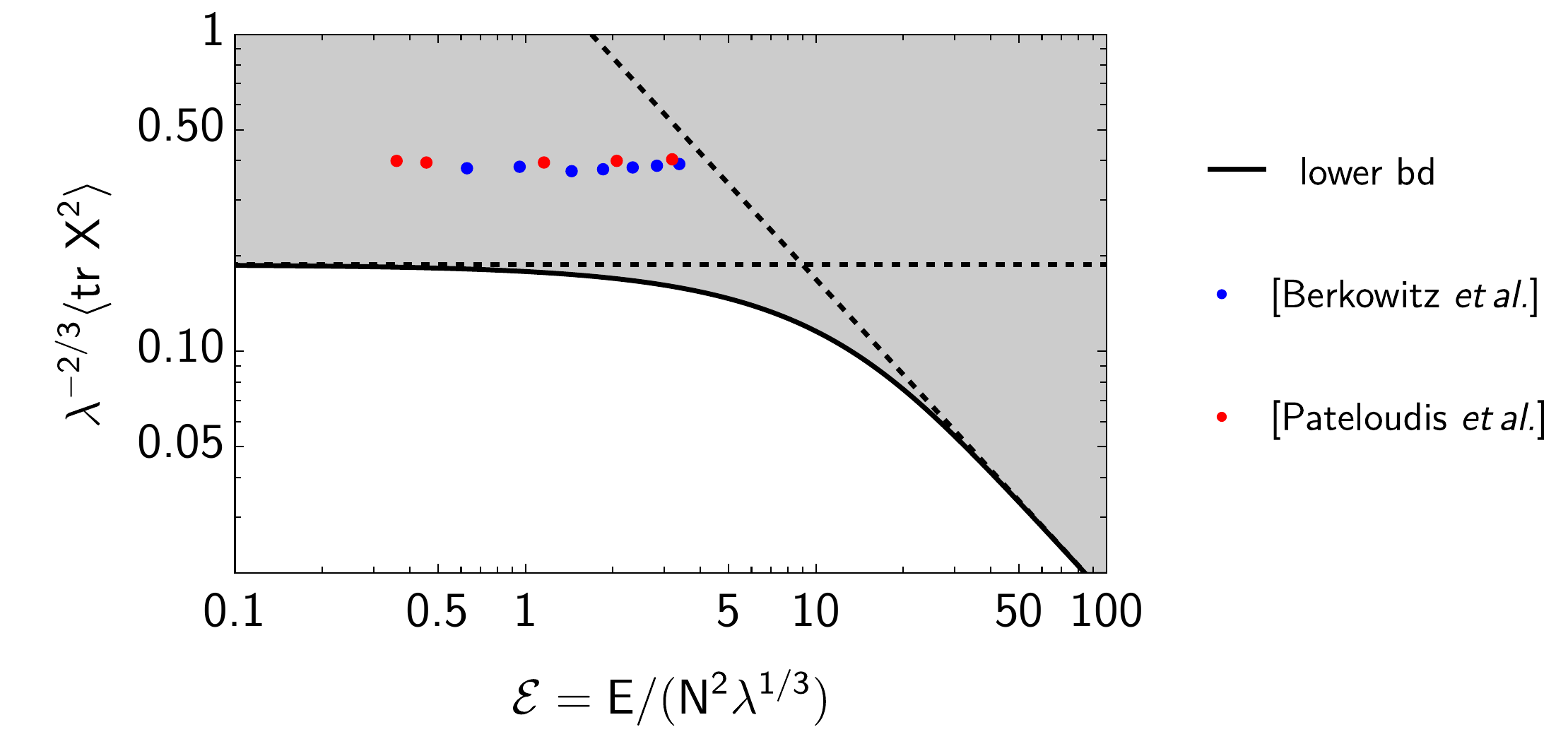}
   \vspace{-.5cm}
    \end{center}       
    \caption{Lower bound on $\ev{\tr X^2}$. We also show the small energy and large energy approximations. The {\color{red} red} dots are large $N$ and continuum extrapolations from the Monte Carlo simulations of \cite{Pateloudis:2022ijr}. The statistical uncertainty of these dots is smaller than size of the dots. There is also a systematic error on the red dots coming from the fact that a small BMN mass deformation is turned on; this systematic error is expected to be negligible. We have also shown in {\color {blue} blue} the $N=32$, max number of lattice sites results from Appendix B of \cite{Berkowitz:2016jlq}. }
    \label{fig:t2}
\end{figure}

Using this fermionic information, we can also improve our bound on $\langle \Tr X^4 \rangle$. 
The best bound is obtained by setting $\det \mathcal{M}_2 = 0$ as in \nref{firstLine}, but using $v = 72 \langle \tr \tilde{X}^4 \rangle$, the boundary value allowed by \nref{ineq_comm}:
\def\nin{\tfrac{1}{9}}
\def\thrd{\tfrac{1}{3}}
\eqn{\la{bigEconstr}
&\langle \tr \tilde{X}^4 \rangle \ge  (t_2)^2, \qquad v= 72 (t_2)^2,\\ %
&\left(\frac{\ce}{9}+\frac{v}{3}\right) \left( 12 \sqrt{2v} - \left(\frac{\ce}{9}-\frac{v}{3}\right)^2\right)=54.
}
This constraint is displayed in Figure \ref{fig:t4_constr}. In principle, there is another bound obtained by setting $v$ to the value given in \nref{secondLine}, but we found that this bound is always weaker than the bound \nref{bigEconstr}.

Since \nref{bigEconstr} is a bit complicated, it is worth deriving simpler bounds, by only demanding that a $2\times 2$ submatrix of \nref{Cm2} is positive semi-definite. Demanding positivity of the lower $2\times 2$ submatrix of \nref{Cm2} gives the ``bosonic'' constraint \nref{Polchinski}. Demanding that the upper $2 \times 2$ submatrix of \nref{Cm2} gives constraints just from the fermionic terms: %
\begin{equation}
\begin{split}
\label{}
\langle \tr \tilde{X}^4 \rangle \ge   \left(t_2\right)^2, \quad  0=\det \mathcal{M}^{(2\times 2)}_2=64 t_2 -\frac{4}{729} (\ce - 216 t_2^2)^2. %
\end{split}
\end{equation}
 We see in Figure \ref{fig:t4_constr} that at large energies, the fermionic constraints are more activated whereas at small energies the bosonic constraints play a stronger role. Note however, that even at $E=0$ the Polchinski bound is improved by incorporating the fermionic constraints. Indeed, the Polchinski bound gives $\ev{\tr \tilde{X}^4} \ge 1/16 = 0.0625$ whereas our strengthened bound gives $\langle{\tr \tilde{X}^4} \rangle \ge \lp  \left(7-4 \sqrt{3}\right)/256 \rp^{1/3} \approx 0.06546$.

\begin{figure}
      \begin{center}
  \hspace{1cm}\includegraphics[width=0.7\columnwidth]{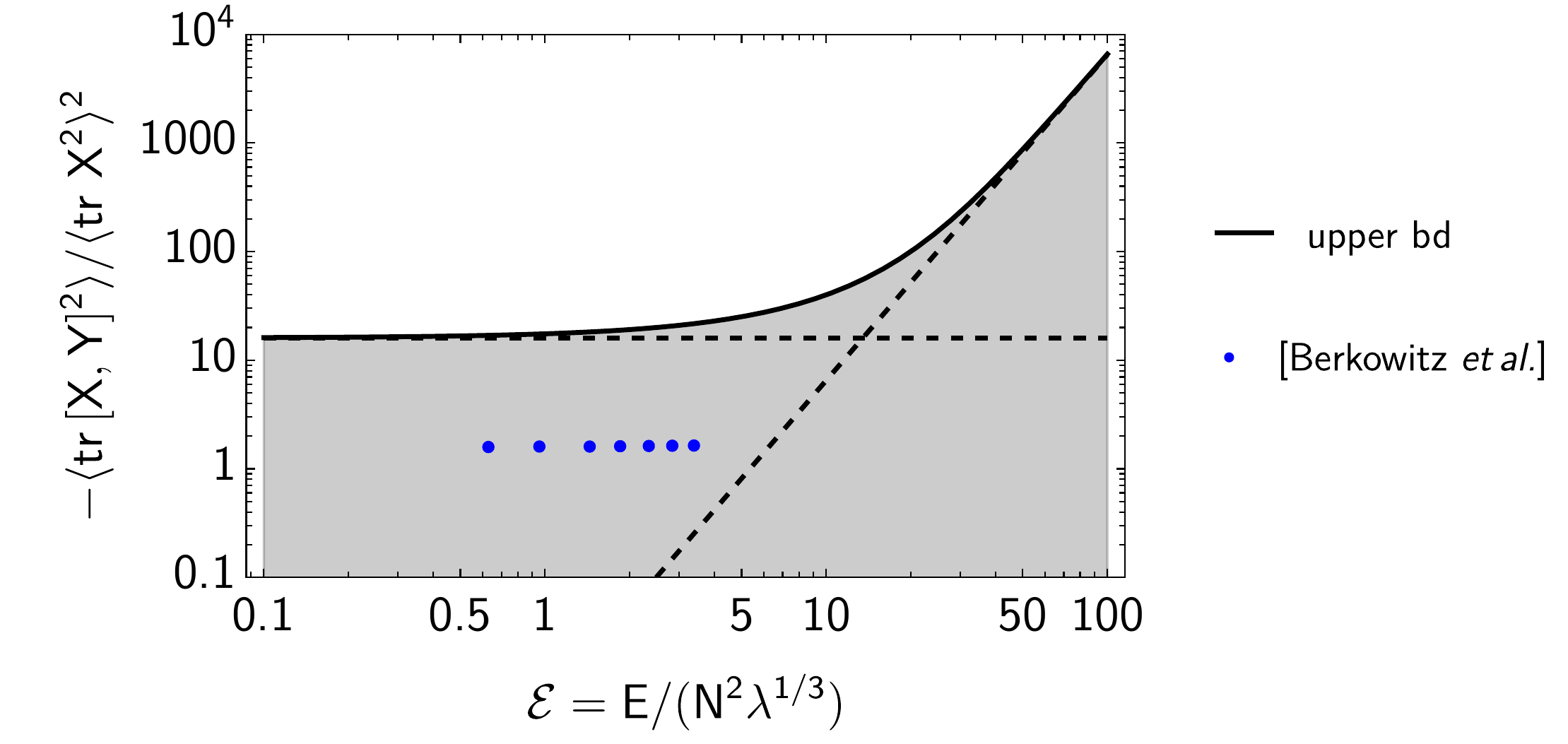}
   \vspace{-.5cm}
    \end{center}       
    \caption{Upper bound on $\gamma = - \langle{\tr [X,Y]^2}\rangle/\langle{\tr X^2}\rangle^2$. The dashed curves are the small energy and large energy approximations. The blue dots are the $N=32$, max number of lattice sites Monte Carlo results from Appendix B of \cite{Berkowitz:2016jlq}. 
    }
    \label{fig:gamma}
\end{figure}

As a final application of the bootstrap, let us consider the quantity 
\begin{equation}
\begin{split}
\label{}
\gamma = \frac{- \langle{ \tr[X,Y]^2} \rangle }{ \ev{\tr X^2}\ev{\tr Y^2} }.
\end{split}
\end{equation}
$\gamma$ quantifies the failure of $X,Y$ to commute, in such a way that is $\gamma$ is invariant under an overall rescaling $X \to a X, \, Y \to b Y$. Reusing once more \nref{firstLine}, we obtain
\begin{align}
\label{gammaBd}
&  0 \le \gamma \le \frac{v_\gamma}{18 } \lb \frac{4(\mathcal{E} - 3 v_\gamma)^2}{9^3 \times 16} + \frac{27}{8(\mathcal{E} +3 v_\gamma)}\rb^{-2},\\ 
& (\mathcal{E} -3 v_\gamma)(\mathcal{E}+3v_\gamma)^2=2 \times  3^9,
\end{align} 
where $v_\gamma$ is defined by minimizing the RHS of \nref{gammaBd}. Since we also have the trivial lower bound $0 \le \gamma $, this is the first case where we obtain a 2-sided bootstrap bound on a quantity. Note that for any ensemble $\rho$ that is invariant under SO(9) rotations, knowing $\ev{\Tr X^4}$ and $\ev{\Tr[X,Y]^2}$ is in fact equivalent to knowing all correlators that are quartic in the $X^I$'s, e.g, $\ev{\Tr X^I X^J X^K X^L}$ since there are only two quartic operators that are SO(9) singlets, see Appendix \ref{SO9}. 
\section{Discussion}
\begin{enumerate}
	\item The constraints we derived did not use large $N$ in any important way\footnote{We dropped some factors of $1/N$ for simplicity at the end of Appendix \ref{fermionMatrices}, but these could be easily restored. These factors strengthen the bounds at finite $N$ by a very slight amount.}. Our bounds also apply at finite $N$, away from the 't Hooft limit. Like Polchinski's virial bound, they apply at very low energies, where the dual is in principle 11D M-theory. Of course, to probe the M-theory region, one would presumably need to compute more sophisticated observables, see also the interesting discussions in \cite{Azeyanagi:2012xj, Bergner:2021goh}. We expect that when further constraints are derived, it will be important to impose large $N$ factorization, which will enforce that we are in the 't Hooft regime.
	\item Another unused input is the SU(N) gauge constraint. So our bounds apply to both the gauge and ungauged model \cite{Maldacena:2018vsr}. At low energies, we do not expect the gauge constraint to play an important role. It would be interesting to bootstrap the minimum energy of a gauge non-singlet representation $R$; according to \cite{Maldacena:2018vsr} this energy should be of order $E \sim \lambda^{1/3}$. To input this into the bootstrap, one would impose that the Casimir formed by the SU(N) generators is given by the value that is set by the representation $R$ of interest.

	 	\item Despite using the structure of the fermion terms in the Hamiltonian, we did not make explicit use of supersymmetry. At finite energies, constraints from SUSY are not expected to be particularly powerful, but perhaps they are useful to bootstrap properties of the ground state. If one can compute some observables via localization \cite{Asano:2014vba}, these computations could be fed into the bootstrap.  An argument due to Susskind \cite{Polchinski:1999br} is that $\ev{\Tr X^\ell}$ for $\ell \ge 9$ does not exist (e.g. diverges), see \cite{Lin:2014wka} for a recent approach. Proving that no finite value of $\ev{\Tr X^\ell}$ is consistent in the ground state would be interesting for the bootstrap.

	\item A quantitative holographic description of $\ev{\Tr X^\ell}$ at various energies would give a target for both Monte Carlo and bootstrap studies, see Appendix \ref{thermal1pt}. Related to this, a concrete prediction from supergravity is the thermodynamics of the model, based on the Bekenstein-Hawking entropy formula. This has already been verified to increasing levels of precision using Monte Carlo \cite{Berkowitz:2016jlq, Pateloudis:2022ijr}. To make contact with this using the bootstrap, one could compute the 1-pt of the Lagrangian in the microcanonical ensemble. This would determine the large $N$ free energy $F(E)$. A long term goal would be to use the bootstrap to obtain high-precision estimates of the stringy $\alpha'$ corrections to various quantities with gravity duals.

\item 

An important problem that we leave to the future is to bootstrap time-dependent correlation functions; see Appendix \ref{SQM} and also \cite{Nancarrow:2022wdr} for a fascinating approach using a quantum mechanical analog of the crossing equation. For example, one would like to consider the connected contribution to the thermal two-pt function $ G_\beta(\tau) = \ev{\Tr M(0) \Tr M(t)}$ . This is conceptually equivalent to considering off-diagonal elements $\bra{E} \Tr M \ket{E'}$. 
A hope is that one could match to gravity expectations, like the quasi-normal modes in \cite{MaldaPrep}. %

\end{enumerate}

We view the simple constraints derived here as a ``proof-of-principle'' that the bootstrap can shed light on the D0 brane system and its black hole dual, despite the complexity of the quantum system. Undoubtedly one can derive more bounds on the BFSS or the BMN matrix model. To efficiently automate this exercise, one should leverage the symmetries of the problem. Presumably one would like to organize the operators into representations of the SO(9) symmetry, or multiplets of the supersymmetry algebra.  Now that a lower bound on the power of the BFSS bootstrap has been provided, the search should continue until an upper bound is achieved.

\section*{Acknowledgements}
I thank Adam Brown, Yiming Chen, Masanori Hanada, Himanshu Khanchandani, Juan Maldacena, Victor Rodriguez, Stephen Shenker, and Xi Yin for discussions. Particular thanks to Stratos Pateloudis and the authors of \cite{Pateloudis:2022ijr} for sharing some of their Monte Carlo data.

I am supported financially by a Bloch Fellowship.

\appendix

\section{Bound on $\Tr X^2$ using the supergravity solution \la{thermal1pt}}
The BMN model is a deformation of the original model by adding the following terms to the Euclidean action:
\begin{equation}
\la{bmn_action}
\begin{aligned}
I_\mu=\frac{N}{2\lambda} \int \diff t & \operatorname{Tr}\left[\left(\frac{\mu}{3}\right)^2 \sum_{a=1}^3\left(X^a\right)^2+\left(\frac{\mu}{6}\right)^2 \sum_{i=4}^9\left(X^i\right)^2+\frac{\mu}{4} \psi \gamma_{123} \psi+ \i \frac{2  \mu}{3} \sum_{a, b, c=1}^3 \! \!X^a X^b X^c \epsilon_{a b c}\right].
\end{aligned}
\end{equation}
The idea is that by differentiating the thermal partition function twice, we should get an estimate of the $\Tr X^2$. More precisely,
\begin{equation}
\begin{split}
\label{partition_d}
 \pd_\mu^2 \log Z(\beta,\mu)  \big|_{\mu =0} %
&=-\frac{ N  \beta}{2\lambda}   \ev{\Tr X^2} +\ev{O_3 O_3} -\ev{O_3}^2,\\
O_3 &= {N \over \lambda} \lp \frac{1}{8} \psi \gamma_{123} \psi +\frac{\i }{3} X^a X^b X^c \epsilon_{abc} \rp .
\end{split}
\end{equation}
In the last line, we used that at $\mu= 0$ the full SO(9) rotational invariance is restored, so we could replace $X^I$ with any one of the bosonic matrices $X$. 
Note that due to the factor of $N$, we cannot ignore the connected correlator of $O_3$, even in the 't Hooft limit. However, since this term is a variance, we know it is positive, so we can use this to get a bound on $\ev{\Tr X^2}$.

Now the idea is that $Z(\beta,\mu)$ can be computed using the gravity solution \cite{Costa:2014wya}; see their equation (71) and their Figure 9. The solution is rather complicated but the free energy can be computed numerically. For $\mu/T \ll 1$, 
\begin{equation}
\begin{split}
\label{}
\log Z =  N^2 \lp 4.11 (T/\lambda^{1/3})^{9/5}  \rp \lp 1 - 0.33 \hat{\mu}^2 + \cdots \rp, \quad \hat{\mu} = 7 \mu/(12 \pi T).
\end{split}
\end{equation}
So plugging this free energy into \nref{partition_d}, we get
\begin{equation}
\begin{split}
\label{}
\lambda^{-2/3}  \ev{\tr X^2} \ge k    \lp T/\lambda^{1/3} \rp^{4/5} \approx 0.11\mathcal{E}^{4/14}, \quad k \approx 0.19.
\end{split}
\end{equation}
Here we converted to energy using \nref{thermo}. 
Note that strictly speaking this bound is only valid when $T/\lambda^{1/3} \ll 1$, where it is quite weak. At higher temperatures, one should include $\alpha'$ corrections to the free energy. At $T=0.35$ where the Monte Carlo results agree with supergravity, this gives a lower bound that is $\sim 20\%$ of the Monte Carlo result for $\ev{\Tr X^2}$.

It would be nice to estimate the 1-pt function of other operators like $\Tr X^n$ using the supergravity solution. The holographic interpretation of operators $\Tr X^n$ is not well understood, but we can consider a related problem\footnote{I thank Juan Maldacena for suggesting this.} of computing the 1-pt function of a massive string mode $\Phi$ which couples to the Weyl curvature $\sim \alpha \Phi W^2$ \cite{Grinberg:2020fdj}. Using \nref{metricIIA}, we may compute the Weyl curvature scalar. At large $r$, $W^2 \sim  1/R_\text{eff}^{4}$ where $R_\text{eff}$ is the effective radius of the $S_8$. It therefore seems plausible that one can interpret the 1-pt function as coming from primarily the large curvature region; for a quantitative understanding, one would need to know how to regulate the large $r$ divergence.

As a general remark, let us comment on how to use the bootstrap to compute thermodynamics at large $N$. This would enable a test of the Bekenstein-Hawking formula. %
The bootstrap is good at computing 1-pt functions in the microcanonical ensemble. %
Although the entropy is not a 1-pt function, for thermal ensembles at large $N$, there is a simple trick. 
One simply computes the 1-pt function of the Euclidean Lagrangian in this theory $\mathcal{L}$. This is a single-trace operator, so in principle we can compute it using the bootstrap. (In some cases, it might be more efficient to take derivatives with respect to the couplings.)
At large $N$, we expect that the canonical ensemble with temperature $\beta_E$ corresponding to the microcanonical energy $E$ will reproduce the same expectation value for $\ev{\mathcal L}$. But in the canonical ensemble, $- \beta_E \ev{\mathcal{L} } = -\beta_E F$, so $\ev{\mathcal{L}}$ has the interpretation of the free energy $F(\beta_E) = F(E)$. Hence we may determine the large $N$ thermodynamics. (These comments generalize to include chemical potentials.) So solving the large $N$ thermodynamics does not require any new bootstrap technology (at least as a matter of principle) beyond what has already been developed.

\section{Bootstrapping off-diagonal elements of single particle quantum mechanics}\la{SQM}
\def\eep#1{ \left( #1 \right)_{E,E'} }
\def\eeptr#1{ \left( \Tr #1 \right)_{E,E'} }
\def\eepn#1{ {#1}_{E,E'} }
\def\co{\mathcal{O}}
\def\cre{a^\dagger}

Here we review the bootstrap method of \cite{Han:2020bkb} and generalize it to constrain off-diagonal matrix elements. To explain the generalization, it is useful to consider the simplest possible case, single particle quantum mechanics. \cite{Han:2020bkb} considered bootstrapping the expectation values of operators $\{ O_i \} $ in some energy eigenstate $E$. The idea was that one can leverage 
\eqn{\mel{E}{[H,O_i]}{E} = 0 ,}
by using that $[H,O_i]$ generates relations between higher-pt correlators and lower-pt correlators. One combines this with positivity 
\eqn{\evv{\mathcal{O}^\dagger \mathcal{O}}{E} \ge 0.} 
In the bootstrap approach, the unknown parameters are $E$ together with some set of correlations functions. One then derives constraints on these values. For each value of $E$, we get some possible range in the correlators that in favorable situations shrinks to zero. 
The new idea is that we can also bootstrap matrix elements. We use the identity 
\eqn{\mel{E}{[H,O_i]}{E'} =(E-E') \mel{E}{O_i}{E'}}
together with positivity 
\eqn{ \evv{\mathcal{O}^\dagger \mathcal{O}}{\psi} \ge 0 , \quad \ket{\psi} = c \ket{E} + d \ket{E'} .} 
(In principle, we could also consider more positivity constraints if we are interested in multiple values of $E, E'$.) This is essentially the constraint
\begin{equation}
\begin{split}
\label{}
  \evv{\co^\dagger \co }{E} \evv{\co^\dagger \co }{E'} \ge |\mel{E}{\co^\dagger \co}{E'}|^2, \quad   \evv{\co^\dagger \co }{E} \ge 0.
\end{split}
\end{equation}
Notice that the quantity that is naturally constrained is the frequency-space correlator. 
The idea is then that the unknown parameters are $E,E'$ together with some set of correlation functions. %
More concretely, consider a non-relativistic particle in a potential:
\begin{equation}
\begin{split}
\label{asho}
  H = \hf p^2 + V(q) .%
\end{split}
\end{equation}
\def\eps{\Delta E}
Let's adopt the shorthand $O_{E,E'} \equiv \bra{E} O \ket{E'}$ and $\eps \equiv E-E'$. Then we have
\begin{equation}
\begin{split}
\label{asho4}
& -is\eep{ q^{s-1} p} -\frac{1}{2} s(s-1)\eep{q^{s-2}} =(E-E') \eep{q^s} \\
& -im \eep{ q^{m-1} p^2} + i \eep{ q^m V'(q)} - \hf  m(m-1) \eep{q^{m-2}p} =  (E-E') \eep{q^m p} \\
& E'\eep{q^{n-1}}  = \hf \eep{q^{n-1} p^2} + \eep{q^{n-1} V(q)}.\\
\end{split}
\end{equation}
In the last line, we have used $\eep{OH} = E'\eepn{O}$. Now plugging in the first equation into the second equation (twice) to eliminate terms linear in $p$, and then eliminating the $p^2$ term using the third equation gives
\begin{equation}
\begin{split}
\label{recurse}
&  2m\eep{q^{m-1} V(q) } + \eep{ q^m V'(q)} - 2m E'\eep{q^{m-1}} - \qrt m(m-1) (m-2) \eep{q^{m-3}}   \\
&= \eps  \lb {\eps \over m+1}  \eep{q^{m+1}}+ m \eep {q^{m-1}}  \rb. \\
\end{split}
\end{equation}
We can consider $V=\hf q^2 + \qrt g q^4$. Then we get a recursion relation involving the moments $m+3, m+1,m-1,m-3$.
\begin{equation}
\begin{split}
\label{recurseSHO}
&   {g\over 2} (m+2)  \eep{q^{m+3} } = -(m+1)\eep{q^{m+1}} + 2m E'\eep{q^{m-1}} + \qrt m(m-1) (m-2) \eep{q^{m-3}}   \\
& +\eps  \lb {\eps \over m+1}  \eep{q^{m+1}}+ m \eep {q^{m-1}}  \rb. \\
\end{split}
\end{equation}
The bootstrap constraints for this problem are just 
\begin{equation}
\begin{split}
\label{constr}
  \mathcal{M} \succeq 0, \quad \mathcal{M}_{i,j,E_i, E_j} = \bra{E_i} q^{i+j} \ket{E_j}.
\end{split}
\end{equation}
For a discrete spectrum, we input that $\eep{q^0} = \delta_{E,E'}$.

When we write the positivity constraint $\mathcal{M} \succeq 0 $, we are viewing the multi-indices $\{i,E_i\}$ and $\{j, E_j\}$ as the indices of a large matrix; the eigenvalues of this matrix are non-negative.
We can then bootstrap this model as follows. First, we set $\eps =0$ and we restrict the matrix $\mathcal{M}$ to $E_1 = E_2$. This reduces the problem to what was already considered in \cite{Han:2020bkb}, see also \cite{Berenstein:2021dyf, Berenstein:2022unr}.

By maximizing the minimum eigenvalue of the matrix $\cal{M}$, one can obtain estimates of $\{E_,\ev{q^2}\}$. We found that there is a local maximum near the allowed energy eigenvalues of the anharmonic oscillator. One can check on a laptop using {\it Mathematica}'s arbitrary precision arithmetic\footnote{A simple method to compute the eigenvalues of the anharmonic oscillator to high precision is to consider the Hilbert space of an oscillator, truncated to level $N \ge N_c$. We then computed the eigenvalues of \nref{asho} $H = \cre a + \hf + \frac{g}{16} \lp a+\cre \rp^4$ to high precision.} that the first few eigenvalues are reproduced to an accuracy of $\sim 10^{-10}$.

With these eigenvalues in hand, we can then apply the bootstrap to the off-diagonal elements. The results from imposing a modest number of constraints $m \sim 14$ are displayed in Table \ref{table}. A small nuisance is that the bootstrap only gives upper bounds on the absolute values of the matrix elements. The reason is that the overall phase of the matrix element is unphysical since it can be redefined by rotating $\ket{E_i} \to e^{-i  \theta_i} \ket{E_i}$. Nevertheless, we see that the upper bounds come very close to saturating the answers from exact diagonalization.
\begin{table}[hbt]
\begin{center}
  \begin{tabular}{c|c|c|c}
 & diagonalization   &  lower bd  & upper bd  \\
    \hline
 $\evv{q^2}{E_1}$   		& 0.354840 & 0.354878 & 0.354828 \\
  $\evv{q^2}{E_3}$   		& 1.395413  & 1.395390 & 1.395423 \\
    $\evv{q^2}{E_5}$   		& 2.142936  & 2.142922 & 2.142942 \\
$|\mel{E_1}{q^2}{E_3}|$    & 0.4769697  & 0 & 0.4769697\\
$|\mel{E_3}{q^2}{E_5}|$    &  0.94379722 & 0 & 0.94379724\\
$|\mel{E_1}{q^2}{E_5}|$    & 0.036958531  & 0 & 0.036958551\\
  \end{tabular}
  \caption{SDP Bounds on matrix elements of $q^2$ for $g=1$. \la{table} }
\end{center}
\end{table}

\section{Fermionic matrices \la{fermionMatrices} }
The fermionic matrices are defined by
\begin{equation}
\begin{split}
\label{majorana}
  \psi_\alpha = \psi_\alpha^A \t_A, \quad  \Tr( \t^A \t^B) = \delta^{AB}, \quad  \{ \psi_\alpha^A, \psi_\beta^B\} =\delta^{AB} \delta_{\alpha \beta}.
\end{split}
\end{equation}
Here $\t^A$ are generators of su($N$). (We have removed the degrees of freedom corresponding to the center of mass.)
With these conventions,
\begin{equation}
\begin{split}
\label{ttCas}
  \t^A \t^A = \frac{N^2-1}{N}, \quad \Tr \t^A \t^B \t^A \t^C = - \frac{1}{N}\delta^{BC} 
\end{split}
\end{equation}
\begin{equation}
\begin{split}
\label{psi2}
  \Tr \Psi^2 = \Psi^A \Psi^B \Tr(\t^A \t^B) = \Psi^A \Psi^A  = \hf (N^2 - 1). \\
\end{split}
\end{equation}
Using these identities we may compute $\Tr \Psi^4$:
\begin{equation}
\begin{split}
\label{psi4}
  \Tr \Psi^4 &= \Psi^A \Psi^B \Psi^C \Psi^D \Tr \t^A \t^B \t^C \t^D\\
   &= -\Psi^B \Psi^A \Psi^C \Psi^D  \Tr \t^A \t^B \t^C \t^D  + \Psi^C \Psi^D \tfrac{N^2-1}{N} \Tr \t^C \t^D\\
  &= \Psi^B  \Psi^C \Psi^A \Psi^D  \Tr \t^A \t^B \t^C \t^D  - \Psi^B \Psi^D \Tr \t^A \t^B \t^A \t^D+  \tfrac{(N^2-1)^2}{2N}  \\
  &= -\Psi^B  \Psi^C  \Psi^D \Psi^A  \Tr \t^A \t^B \t^C \t^D  +  \frac{(N^2-1)^2}{N} + \frac{N^2-1}{2N},  \\
   \RA \Tr \Psi^4 &= \frac{1}{2N} \lp N^4 - \frac{3N^2}{2} +\hf \rp  < \hf N^3.
\end{split}
\end{equation}
Here we have used both equations in \nref{ttCas}. %
Now let's consider bounding $\Tr \Psi^2  \Phi^2$ where $\Psi, \Phi$ are different fermionic matrices. Using \nref{ttCas} and \nref{psi2}, we obtain:
\begin{equation}
\begin{split}
\label{posAB}
 \Psi^A \Psi^B \Phi^C \Phi^D \Tr \t^A \t^B \t^C \t^D &= (- \Psi^B \Psi^A \Phi^C\Phi^D + \delta^{AB} \Phi^C\Phi^D) \Tr \t^A \t^B \t^C \t^D\\
 &= -\Psi^A \Phi^B \Phi^C \Psi^D  \Tr \t^A \t^B \t^C \t^D + (N^2-1)^2/(2N^2)\\
 &= - \Tr (\Psi \Phi) (\Psi \Phi)^\dagger + (N^2-1)^2/(2N^2) <  1/2.
\end{split}
\end{equation}
We can consider the $(1+16) \times (1+16)$ inner product matrix $\mathcal{N}$ of correlators:
\begin{equation}
\begin{split}
\label{MAB}
\mathcal{N} = \begin{pmatrix}
	1 & N^{-2} \Tr (\psi_\alpha)^2  \\
	 N^{-2} \Tr (\psi_\alpha)^2 &   \mathcal{M}_{\alpha \beta} \end{pmatrix}, \quad 
  \mathcal{M}_{\alpha \beta} = \frac{1}{N^3} \ev{\Tr \psi_\alpha^2 \psi_\beta^2 }, \
\end{split}
\end{equation}
Using \nref{psi2}, we may infer the value of the first row and column of $\mathcal{N}$, and using  \nref{psi4}, we learn the values of the diagonal. Furthermore, \nref{posAB} implies $\mathcal{M}_{\alpha \beta} < \hf $ for $\alpha \ne \beta$.  Then we may consider the semi-definite problem of maximizing $\sum_{\alpha, \beta} s_\alpha s_\beta \mathcal{M}_{\alpha \beta}$ with these constraints $\mathcal{N} \succeq 0$. We find:
\begin{equation}
\begin{split}
\label{result4}
  \sum_{\alpha, \beta} s_\alpha s_\beta \mathcal{M}_{\alpha \beta} \le 64(1-\epsilon') \le 64.
\end{split}
\end{equation}
Here $\epsilon'$ is a positive number that approaches 0 in the large $N$ limit, so we drop it. %

\section{Rotational invariance \la{SO9}}
In this appendix, we comment on how to leverage the SO(9) rotational invariance for the bootstrap.
If the ensemble we are interested in is also rotationally invariant (in the sense that $[J^{IJ}, \rho]= 0$), the only non-zero observables are SO(9) singlets. However, when we consider positivity constraints, it is useful to consider non-singlet operators.
For example, we may consider the operators
\begin{equation}
\begin{split}
\label{}
  \mathcal{O} = \{ 1,  \, X^I, \, X^I X^J\}.
\end{split}
\end{equation}
We can imagine computing inner product matrix of these operators $\mathcal{M}_{4}$. This is an $L \times L$ matrix, where $L = 1 + 9 + {9 \choose 2}$. The non-zero matrix elements of $\mathcal{M}_4$ are either a constant, quadratic, or quartic in the $X^I$'s.  Imposing SO(9) symmetry, we find that all matrix elements of $\mathcal{M}_{4}$ are determined by just 3 parameters:  
\begin{equation}
\begin{split}
\label{}
\ev{\Tr (X^I X^J)} &= A_2 \delta^{IJ} \\
\ev{\Tr (  X^I X^J X^K X^L)} &= A_4 (\delta^{IJ} \delta^{KL} + \delta^{JK} \delta^{IL}) + B_4\delta^{IK} \delta^{JL}.
\end{split}
\end{equation}
On the RHS we have listed various tensors that correspond to different ways to make an SO(9) singlet. We have also imposed cyclicity of the trace. 
Note that $\Tr X^4 =2A_4+B_4$ and $\Tr [X,Y]^2 = 2(B_4 -  A_4)$. Then demanding that $\mathcal{M}_4 \succeq 0$, we may derive various inequalities on $A_4,B_4$ and$A_2$. For example,
\begin{equation}
\begin{split}
\label{strongerSO9}
  \ev{\tr X^4} \ge \text{max}\left\{-\frac14 \langle{\tr [X,Y]^2}\rangle , \,\, \frac{4}{11} \langle{\tr [X,Y]^2}\rangle + \frac{27}{11}  \ev{\tr X^2}^2  \right\}.
\end{split}
\end{equation}
The second inequality is interesting because for an $SO(D)$ symmetry, the constants that appear are $D$-dependent (here we used $D=9$). Note in particular that the weaker constraint $\ev{\tr X^4} \ge \ev{\tr X^2}^2$ can be derived from \nref{strongerSO9} by looking at the point when the two terms in the maximum are equal, e.g., when $-\frac{27}{44}\tr [X,Y]^2 = \frac{27}{11} \ev{\tr X^2}^2$.  For the simple constraints derived in, e.g., Figure \ref{fig:t4_constr}, we found that inputting the stronger constraint \nref{strongerSO9} did not lead to any improvement in the bounds.

We could also derive analogous expressions for the fermions, e.g.,
$\ev{\tr \psi_\alpha \psi_\beta \psi_\gamma \psi_\delta} \propto A \gamma^I_{\alpha \beta} \gamma^I_{\gamma \delta} + B \delta_{\alpha \beta} \delta_{\gamma \delta} +  \cdots $. For future purposes bootstrap studies, presumably one would need to consider the general decomposition of a tensor product of vector and spinor representations into singlets.

\bibliographystyle{srt}

\end{document}